# Influence of CO$_2$-laser pulse parameters on 13.5 nm extreme ultraviolet emission features from irradiated liquid tin target

**Vasily S Zakharov**[1-3†], **Xinbing Wang**[1,4*], **Sergey V. Zakharov**[3,5], **Duluo Zuo**[1,4], **Junwu Wang**[1]

[1] Wuhan National Laboratory for Optoelectronics, Huazhong University of Science and Technology, Wuhan, China
[2] Keldysh Institute of Applied Mathematics RAS, Moscow, Russia
[3] NRC 'Kurchatov Institute', Moscow, Russia
[4] Guangdong Intelligent Robotics Institute, Dongguan, China
[5] EATS, Orsay, France
E-mail: †zakharovvas@gmail.com, *xbwang@hust.edu.cn

**Abstract**. Laser-produced plasma (LPP) induced during irradiation of a liquid tin droplet with diameter of 150 µm and 180 µm by CO$_2$ laser pulse with various pulse durations and energies is considered. The two-dimensional radiative magnetohydrodynamic (RMHD) plasma code is used to simulate the emission and plasma dynamics of multicharged ion tin LPP. Results of simulations for various laser pulse durations and 75÷600 mJ pulse energies with Gaussian and experimentally taken temporal profiles are discussed. It is found that if the mass of the target is big enough to provide the plasma flux required (the considered case) a kind of dynamic quasi-stationary plasma flux is formed. In this dynamic quasi-stationary plasma flux, an interlayer of relatively cold tin vapor with mass density of 1÷2 g/cm$^3$ is formed between the liquid tin droplet and low density plasma of the critical layer. Expanding of the tin vapor from the droplet provides the plasma flux to the critical layer. In critical layer the plasma is heated up and expands faster. In the simulation results with spherical liquid tin target, the CE into 2π is of 4% for 30 ns FWHM and just slightly lower - of 3.67% for 240 ns FWHM for equal laser intensities of 14 GW/cm$^2$. This slight decay of the in-band EUV yield with laser pulse duration is conditioned by an increasing of radiation re-absorption by expanding plasma from the target, as more cold plasma is produced with longer pulse. The calculated direction diagrams of in-band EUV emission permit to optimize a collector configuration.

## 1. Introduction

Intensively emitting plasmas of multiply-charged ions (further *multicharged ion plasma*) have been extensively studied in the last decade for radiation sources in the soft X-ray and extreme ultraviolet (EUV). Interest has been stimulated by various possible applications in academic and applied research to study radiation absorption, angular resolved reflectivity, diffraction and scattering processes in matter, metrology, lithography applications etc.

Laser pulses and electric discharge current may be used individually or in concert to heat up and to confine the plasma at a suitable temperature. The harder photon energy with high enough efficiency is required, a higher plasma temperature and laser intensity or pulsed-power should be used. Laser-produced plasma (LPP) is derived through an interaction of focused single or multiple laser pulses (of the same or different wavelengths) with a solid or liquid target [1-2], or gas-puff target [3] in a vacuum chamber; for discharge-produced plasma (DPP), it is an electric discharge in a capillary filled with a gas mixture [4,5] or vacuum arc [6]. The only way to use the discharge technique with solid or liquid targets to obtain ionized plasma is with the combined approach: initially a laser pulse vaporizes and preionizes the target, setting up an arc discharge as the main driver of plasma heating [6-8].

To generate EUV radiation at 13.5 nm wavelength (92 eV photon energy) important for lithography applications, LPP sources using a tin target are the current industry source choices [9–11], where the in-band emission into 2% waveband is resulting from 4d-4f, 4p-4d and 4d-5p transitions from a range of overlapping tin ions $Sn^{6+}$ to $Sn^{13+}$ [12]. To achieve a suitable ionization degree to emit effectively photons with 92 eV energy, the high Z tin plasma should be heated up to temperature of 30-50 eV. Understanding of the physics of non-equilibrium multicharged ion plasma and interaction of the laser light with it is critical for the study of EUV source. Accurate numerical simulation of plasmas together with ionization phenomena and radiation transfer with at least 2-D effects has been recognized as an essential part in the further understanding of plasma behavior and in the development of plasma radiation sources.

Actually, the most developed scheme of LPP EUV source for high-volume manufacturing (HVM), is based on combined irradiation of the mass-limited target (liquid tin droplets with diameter of 20-30μm) with laser pulses of Nd:YAG laser as pre-pulse and of longer wavelength $CO_2$ laser as main pulse at high 50–100 KHz frequency operation [13,14]. The mass-limited droplets are used for the debris reduction at high frequency operation regime. The relatively low-energy short wavelength pre-pulse is used to explode a liquid target and to prepare a large low-density vapour/mist target for the following (in microsecond range delay) longer wavelength main pulse [15]. But a number of laboratories use less expansive and less complicated schemes to study EUV sources with single laser pulse irradiation of larger liquid tin droplets or bulk targets [16] at much lower repetition rate. The physics of these schemes differs significantly from the physics of the combined mass-limited scheme, where the main long-wavelength laser pulse doesn't interact with high density target at all. The capability of bigger targets is not exhausted and studied out yet. By the way, ASML continues target development to have possibility to use droplets with diameter up to 140 μm [13].

Laser-plasma modeling results are investigated here to study physical phenomenon of multicharged ion plasma and EUV emission generation features from a spherical tin target as a function of laser energy and pulse width for the parameters similar to experimental setup in Wuhan National Laboratory for Optoelectronics. The two-dimensional radiative magnetohydrodynamic (RMHD) plasma code Zstar [17] was used to simulate the emission and plasma dynamics of multicharged ion tin LPP. Zstar is a computational code designed to focus specifically in the simulation of a multicharged ion plasma in experimental and industrial facilities. The code Zstar is a further development from the well-proven code ZETA [18,19] and designed to facilitate numerical modeling by non-numerical specialists and do not require knowledge of numerical computation. It is adapted, in particular, to simulate DPP and LPP radiation sources.

Calculations were done for LPP EUV radiation source with Ø150÷180 μm tin droplet under $CO_2$ laser interaction for 90 ns full width half maximum (FWHM) pulse duration and 160÷240 mJ pulse energies with experimentally taken temporal profiles to reproduce the conditions of experimental setup. After initial simulations the Gaussian temporal profile for laser power was used to study the influence of $CO_2$-laser pulse duration on 13.5 nm EUV emission features and on the conversion efficiency. Laser energies considered, varied from 75 mJ to 600 mJ corresponding to laser pulse durations from 30 ns to 240 ns.

## 2. RMHD computational code

The RMHD code Zstar for self-consistent modeling of plasma dynamics together with plasma radiation is designed on the basis of the magnetohydrodynamic (MHD) formalism [20] of multicharged ion plasma in a 2-D axially symmetric geometry with radiation transport. The traditional ideal MHD model has been extended to take into account additional effects relevant to a realistic EUV plasma source as the displacement current to be capable to model the situation, when the conductivity tensor turns to zero, for instance, in not ionized neutral gas or insulator.

Plasma radiation properties, ionization and equation of state (EOS), as well as excitation and ionization rates, and plasma kinetic coefficients are calculated by means of interpolations from a set of tables prepared in pre-processing with the Hartree-Fock-Slater (HFS) model [21] and ionization kinetics [17] in both the optically thick LTE, and the transparent non-LTE limits. The actual non-LTE condition at any instant is modeled by analytical interpolation between these two limits [19]. Non-stationary effects in none-quilibrium plasma, when the ionization degree $\bar{Z}$ depends not only on the plasma density and temperature, but also on the time and on the radiation field, through the ionization kinetics described by level kinetic equations [17,21].The radiation field in the quasi-stationary case is found by integrating the radiation transport equation along the trajectories under cylindrical symmetry conditions [22].

The laser light transport is calculated by means of simplified two-direction transfer model [23] taking into account an absorption and reflection of the laser light along its trajectories. The laser light absorption coefficient includes an interaction of radiation with electrons, ions and neutral atoms: by means of collisional (inverse-bremsstrahlung) absorption with plasma dispersion properties [24], resonant absorption (a critical density effect due to Longmuir plasma oscillation resonant excitation) [24], the effective bond-bond excitation calculated from spectral tables at laser quantum energy with taking into account the induced deexcitation and a probability of dissipation of absorbed laser energy in bond-bond electron excitation to thermal energy, direct (if laser quantum energy is higher than the photo-ionization threshold) or tunnel (if laser quantum energy is lower than the ground state ionization energy) ionization calculated from spectral tables also; the parameter $\gamma$ is introduced to take into account a probability of dissipation of absorbed laser energy in bond-bond electron excitation to thermal energy.

The basic equations underlying the physical model of the code Zstar are presented in Appendix. More detailed description on models, numerical methods and difference schemes, on solvers implemented in Zstar RMHD code, database contents and examples of simulations as well can be found in [17, 25].

## 3. Results and discussion

Extensive simulations were done for LPP extreme ultraviolet (EUV) radiation source with Ø180 µm and Ø150 µm tin droplets under $CO_2$ laser interaction for 30 ns ÷ 240 ns FWHM pulse durations and 75÷600 mJ pulse energies, with gaussian and experimentally taken temporal profiles.

### a. Setup simulation results and laser shape impact study

At the initial stage of simulations, the modeling of LPP EUV radiation source was done with Ø180 µm tin droplet under $CO_2$ laser interaction with 300 µm spot size, 90 ns FWHM pulse duration and 160÷240 mJ pulse energies with experimentally taken temporal profiles to match the experimental conditions [26]. The schematic of the experimental arrangement is illustrated in Figure 1. A self-made tin-droplet generator is used for generating uniform 20 kHz tin-droplets with Ø150-180 µm in $10^{-3}$ Pa vacuum. He–Ne laser is set to probe the tin droplets and serves for a self-made signal delay & syncronization system. $CO_2$ 10.6 µm wavelength laser is employed with 90 ns FWHM, which has a tail of the order of 1-2 µs. It is combined by a selfmade mirror, and the combined laser beam is focused by a commons lens. Laser temporal waveform is acquired by photo electric detector. EUV temporal waveform is recorded by photodiode with a filter composed of Zr films with a thickness of 500 nm positioned in front of the diode.

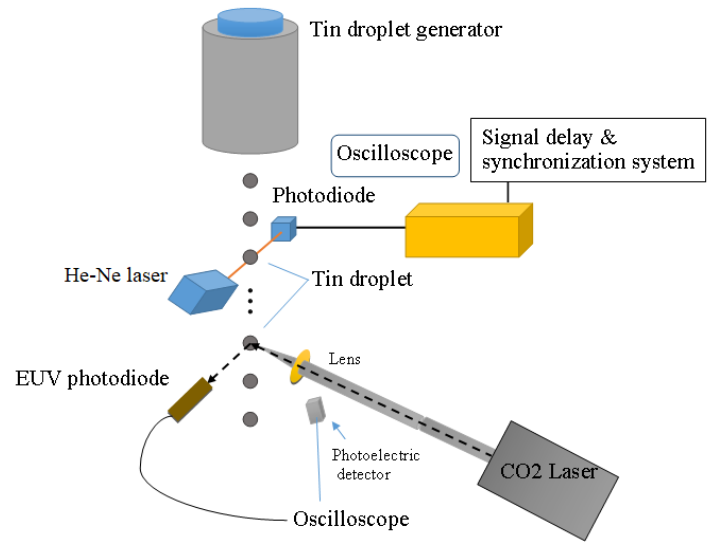

**Figure 1.** Scheme of experimental layout for LPP EUV radiation source in Wuhan National Laboratory for Optoelectronics.

The example of the laser temporal waveform for 240 mJ energy $CO_2$ laser pulse, obtained from experiment and used in the simulation, is shown in Figure 2. The laser pulse starts at about 120 ns and has a long tail over 1.5 µs long, which takes most part of the laser energy – only 88 mJ of 240 mJ total energy is generated in the first 400 ns. In the simulation 55 mJ of these 88 mJ are absorbed by the target and the induced plasma, and the total yield of emitted energy is 24 mJ.

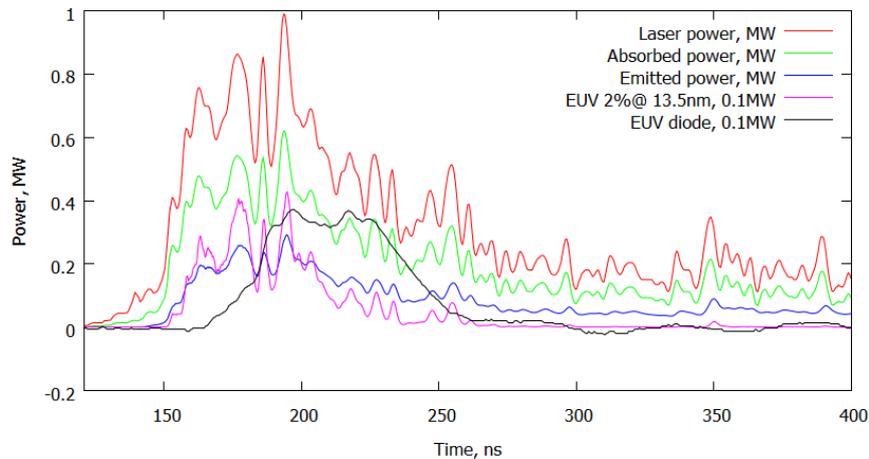

**Figure 2.** Time dependences of laser power (red line), absorbed laser power (green), total emitted power (blue) and emitted power of radiation in the 2% spectral band at 13.5 nm wavelength (magenta, x10 scaled) for laser pulse of 240 mJ energy and corresponding experimentally measured EUV emission power (black, x10 scale factor).

EUV radiation energy output in the 2% spectral band at 13.5 nm wavelength obtained in simulation is of 1.5 mJ. It consists of two strong emission peaks at 177 ns and 195 ns: the power densities at both peaks are shown on Figure 3. On the images, the laser pulse is propagating along the vertical axis and hits on the top of the target. The width of EUV pulse and the amplitude are corresponding well to the data obtained in the experiment by EUV photodiode. EUV signal taken in the measurement is smooth comparing to the computed data – it is explained that Zr filter has a wavelength band wider than 2% bandwidth at 13.5 nm. A delay between the experimental EUV curve and the laser pulse is because of registration specifics in the experimental data treatment.

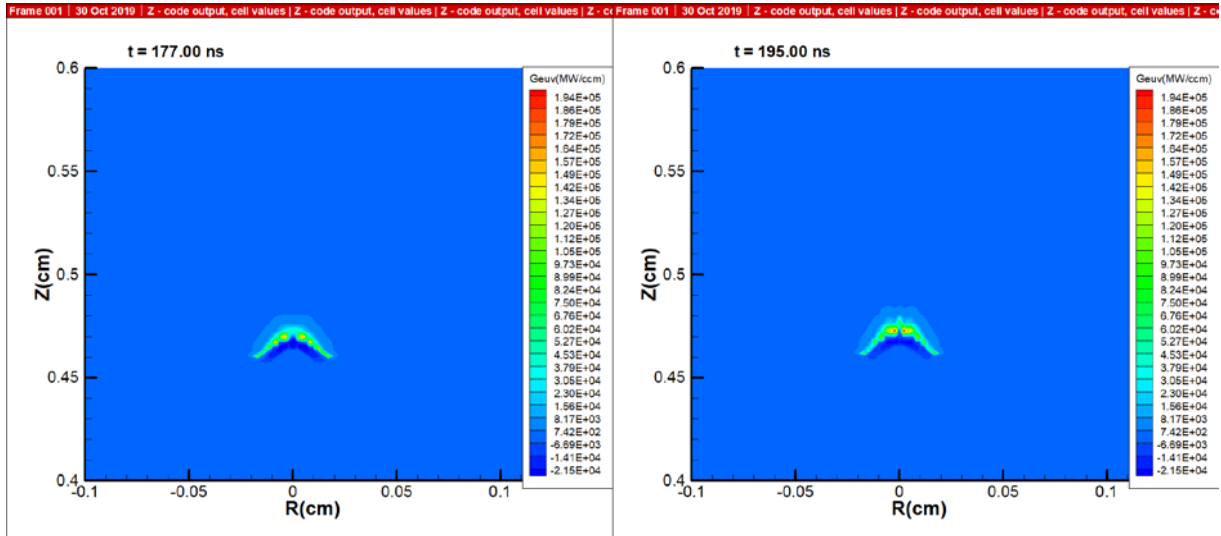

**Figure 3.** Iso-contour cross-sections for in-band EUV emission power density at two EUV emission peaks: at t = 177 ns (left-hand side image) and t = 195 ns (right) for laser pulse of 240 mJ energy.

The temporal evolutions of the plasma mass density, free electron density and electron temperature distribution at various times from 158 to 204 ns are represented in Figure 4. At 177 ns and 195 ns we observe two EUV emission peaks of comparable magnitude. Meanwhile, the corresponding electron temperature spatial distributions at these two instants are rather different: the area with $T_e \geq 30$ eV is less at the first instant than at 195 ns, but the heated area with around two localised spots is larger. It gives a sufficient contribution to EUV generation power, equalizing the emission magnitudes.

The $CO_2$ laser power temporal shape measured in experiment and applied in the simulations discussed above experiences the oscillations of high magnitude. This fact hampers the analysis of obtained results. Strong apparent peaks of EUV emission power can be resulted from either by laser emission peaks influencing on plasma heating and behaviour, or by instabilities in plasma itself. To observe better the dynamics of the plasma and EUV generation phenomenon without such impact from laser pulse variation, another sample of experimental laser power temporal shape was used [27]. The considered laser pulse of 160 mJ energy is smoother than previously discussed and has a less expressed and shorter tail (Figure 5). The pulse has a tail less 1 μs long, 126 mJ of 160 mJ total laser energy is generated in the first 300 ns. According to the simulation results 78 mJ is absorbed by the plasma, and the total yield of emitted energy reaches about 37 mJ. EUV emission power in 2% spectral band at 13.5 nm experiences the oscillations that are much less expressed than in previous case, with the top at about 80 ns time. Iso-contour cross-section for in-band EUV emission power density distribution at this instant is presented in Figure 6. EUV energy output in the 2% spectral band at 13.5 nm wavelength in 4π obtained in simulation is of 3.64 mJ that is more than twice higher of previous case. The temporal evolutions of the plasma mass density, free electron density and electron temperature distribution at various times, and free electron distributions as well are represented in Figure 7 and Figure 8 respectively. It is easy to conclude (see electron temperature distribution, Figure 7 – right-hand side column) that EUV power variations are the result of instability in plasma heating, originating from instability of critical layer. This effect is known and is discussed in detail below in the present work.

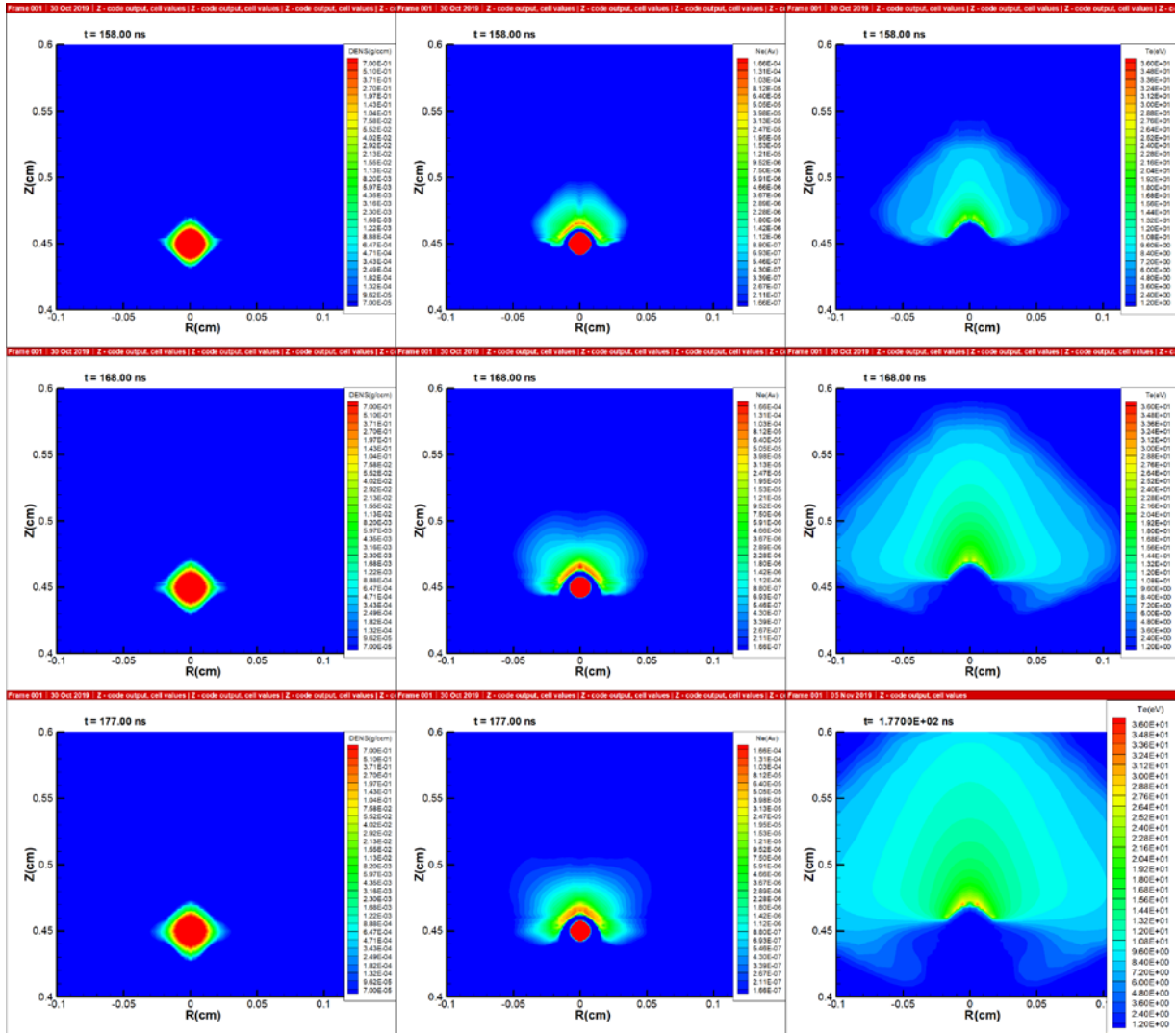

**Figure 4.** Plasma mass density (left-hand side column), free electron density (middle) and electron temperature (right) dynamics as iso-contours of radial cross-section at selected times: at t = 158, 168 and 177 ns. Electron densities are in Avogadro's number units. Images for laser pulse of 240 mJ energy.

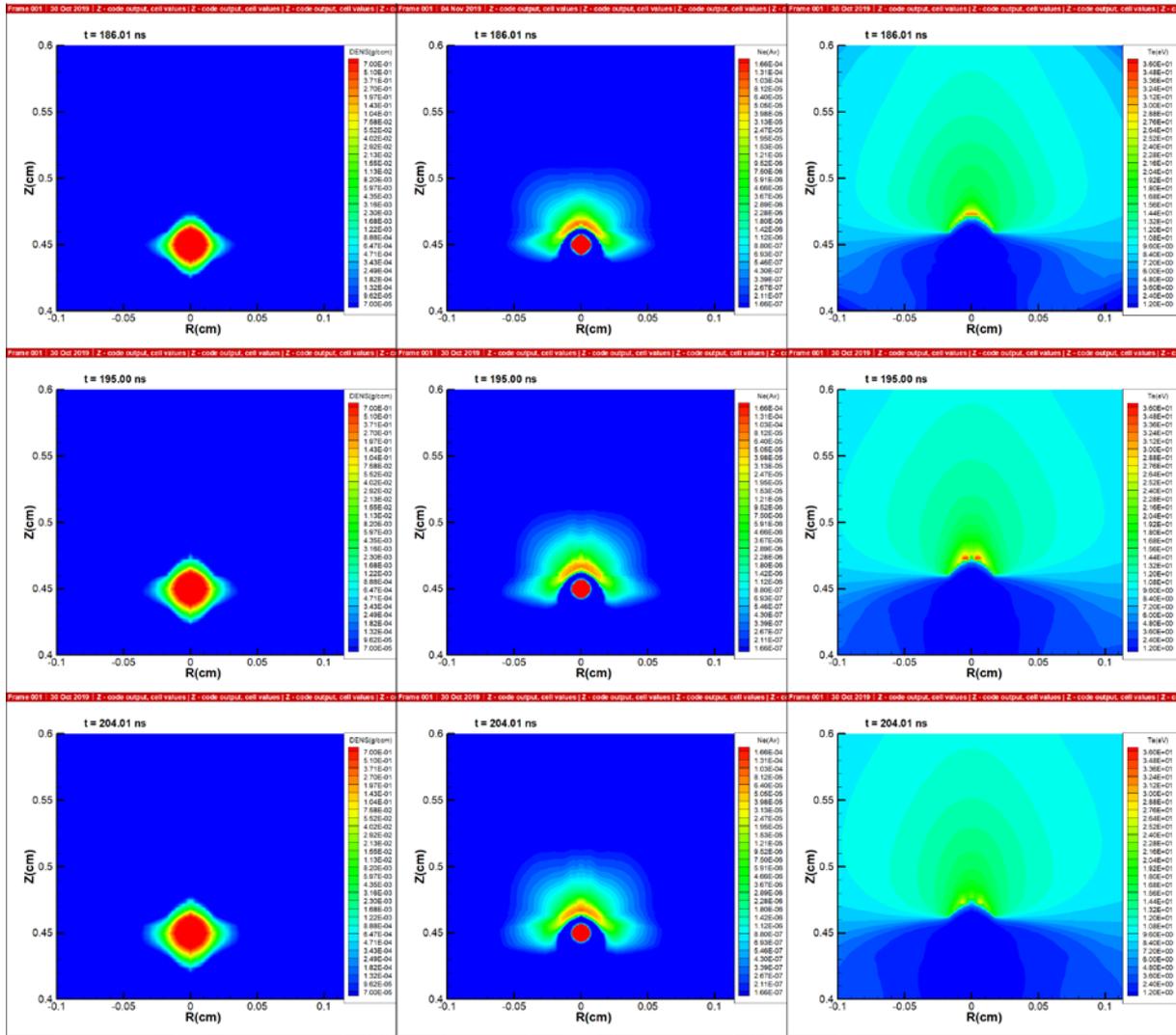

**Figure 4 (continuation).** Plasma mass density (left-hand side column), free electron density (middle) and electron temperature (right) dynamics as iso-contours of radial cross-section at selected times: at t = 186, 195 and 204 ns. Electron densities are in Avogadro's number units. Images for laser pulse of 240 mJ energy.

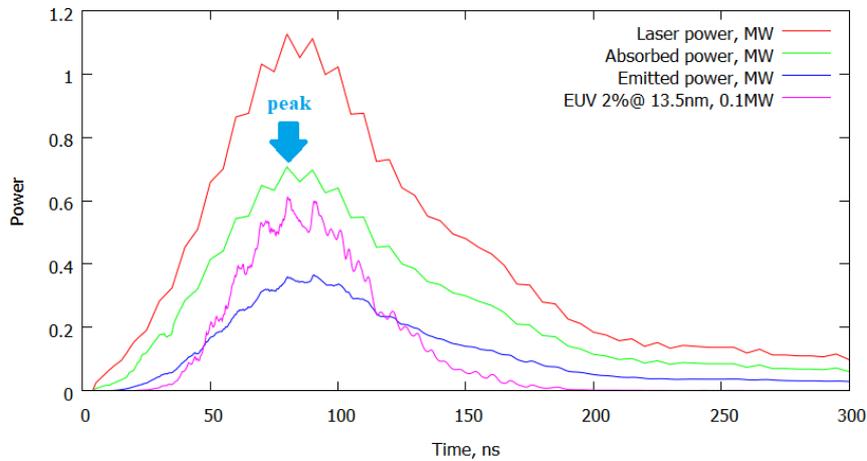

**Figure 5.** Time dependences of laser power (red line), absorbed laser power (green line), total emitted power (blue line) and emitted power of radiation in the 2% spectral band at 13.5 nm wavelength (magenta line, x10 scaled) for laser pulse of 160 mJ energy.

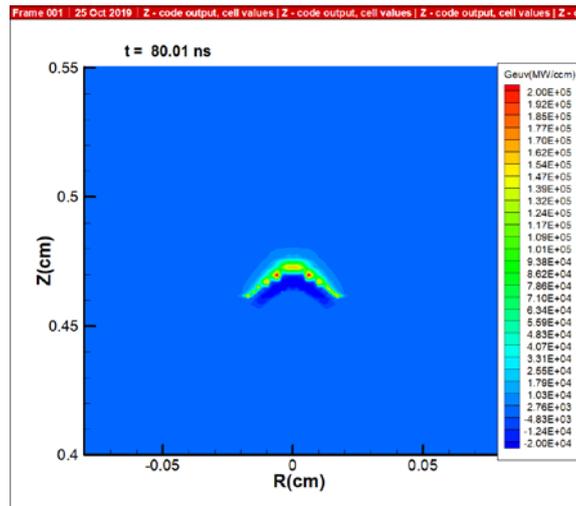

**Figure 6.** Iso-contour cross-section for in-band EUV emission power density distribution at the in-band emission maximum time instant (t = 80 ns) for laser pulse of 160 mJ energy.

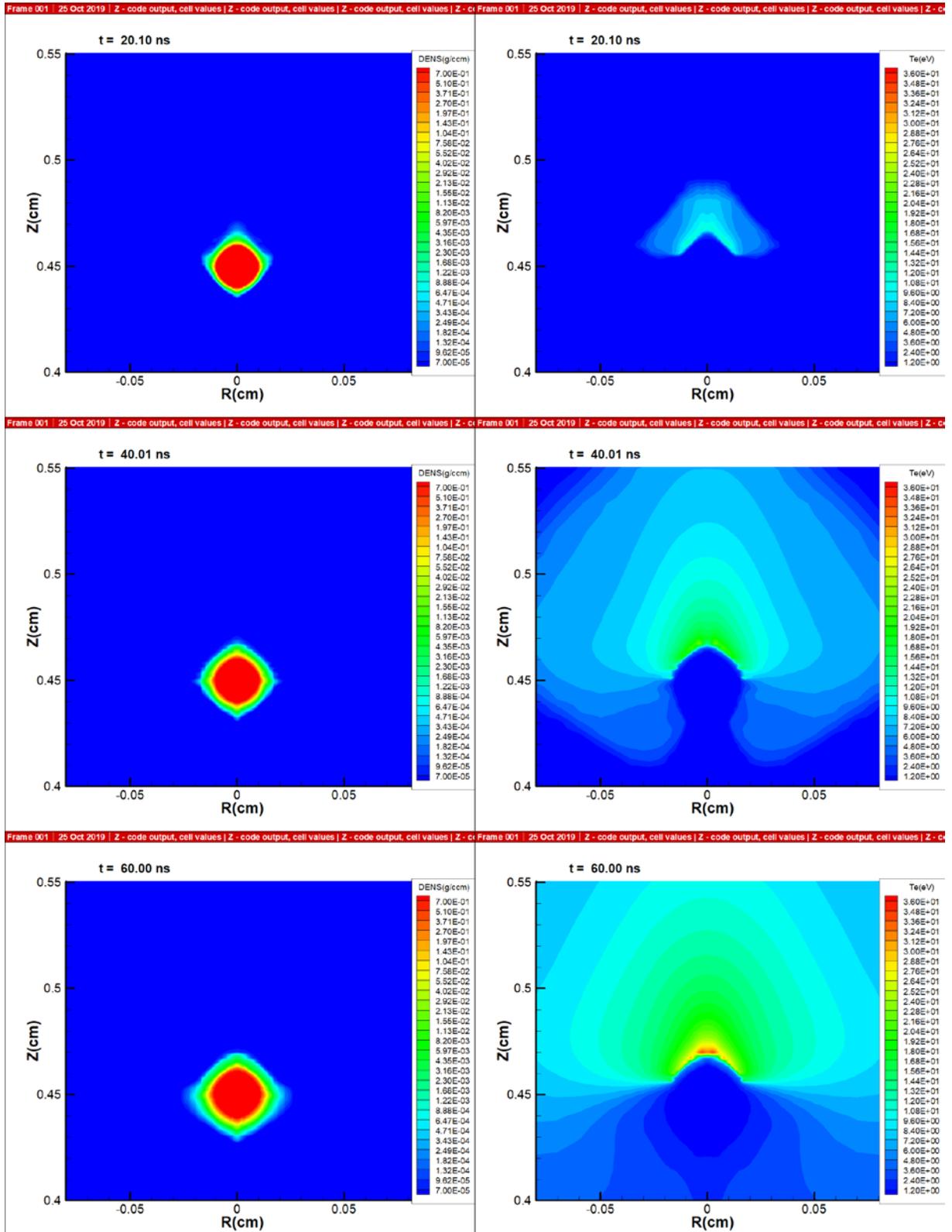

**Figure 7.** Plasma mass density (left-hand side column) and electron temperature (right) dynamics as iso-contours of radial cross-section at various times (from 20 to 60 ns) for laser pulse of 160 mJ energy.

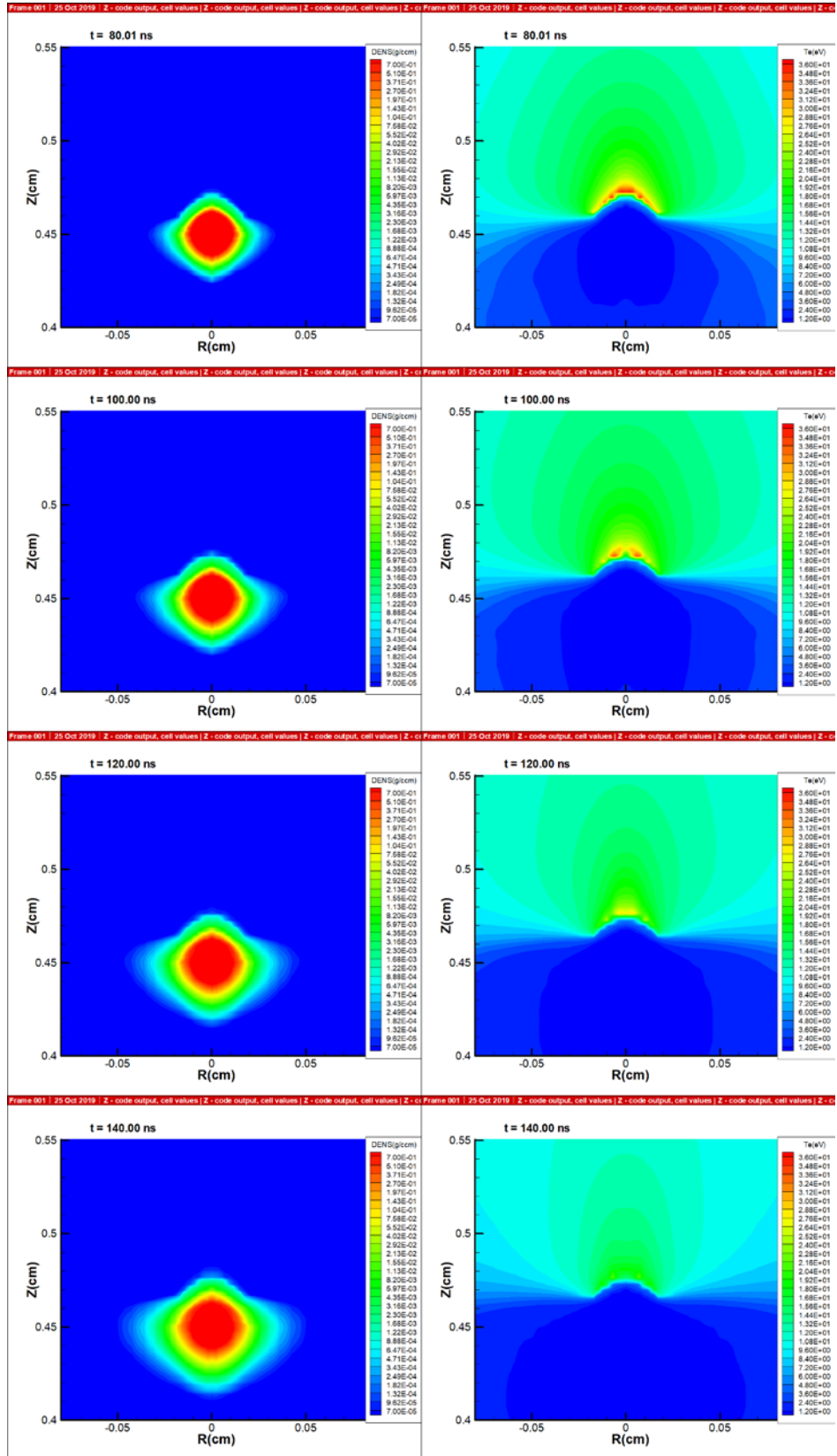

**Figure 7 (continuation).** Plasma mass density (left-hand side column) and electron temperature (right) dynamics as iso-contours of radial cross-section at various times (from 80 to 140 ns) for laser pulse of 160 mJ energy.

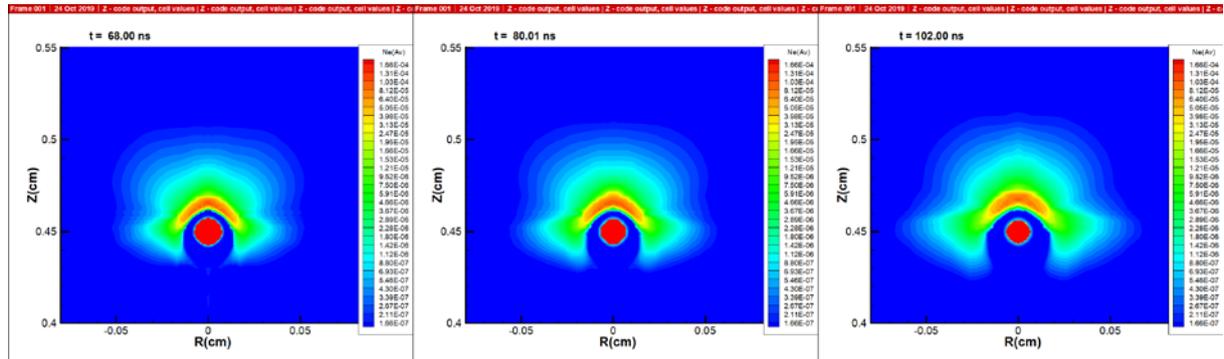

**Figure 8.** Free electron density dynamics as iso-contours of radial cross-section at selected times: at t = 68, 80 and 102 ns. Electron densities are in Avogadro's number units (Av). Images for laser pulse of 160 mJ energy.

**b. Laser pulse duration influence on EUV emission generation**

To investigate the influence of laser pulse duration on plasma behavior and EUV emission generation, a laser-produced plasma (LPP) induced during irradiation of a liquid tin droplet of 150 µm diameter by $CO_2$- laser pulse with various pulse durations at FWHM in the range of 30÷240 ns was considered. Laser intensity amplitude on the target was kept at the level of 14 GW/cm$^2$ and its temporal shape was chosen as Gaussian. The laser beam is focused on target with a focal spot diameter of 150 µm FWHM with Gaussian spatial profile as well. An example of the laser pulse temporal profile together with spectral integrated radiation power and EUV emission in 2% spectral band at 13.5 nm wavelength (so called in-band radiation) is shown in Figure 9 for the laser energy of 150 mJ and the pulse duration of 60 ns FWHM. The laser pulse starts at a time moment of -10 ns, reaches its maximum at 72 ns with power value of 2.35 MW and finishes at 160 ns. The tin plasma generated is heated up and starts to emit in 12÷13 ns after the laser pulse initiation time. The spectrum integrated plasma emission follows the laser pulse mainly and its total yield is of 62.4 mJ. In-band EUV emission is almost ten times lower, it reaches its maximum of 81.5 kW at 67.6 ns, earlier than the laser power has its maximum due to increasing of radiation re-absorption by expanding plasma. After the peak of laser pulse, in the phase of laser power decrease, the hot zone is carried out by the flux to the less dense expanding plasma with lower radiation re-absorption. This explains the local increase of EUV output at 120 ns. The total yield in-band EUV emission is about 5.86 mJ for 150mJ $CO_2$-laser pulse of 60 ns pulse duration.

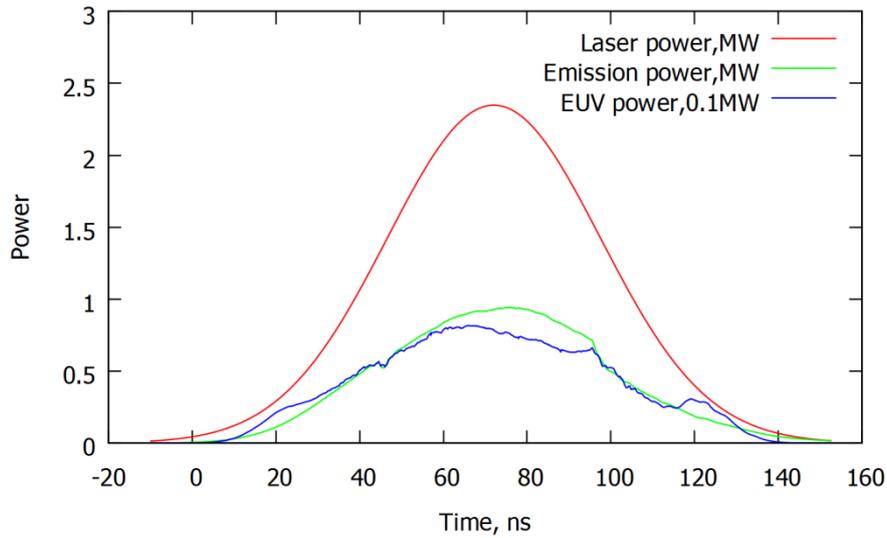

**Figure 9.** Time dependences of laser power (red line), total emitted power (green) and emitted power of radiation in the 2% spectral band at 13.5 nm wavelength (blue, ×10 scaled).

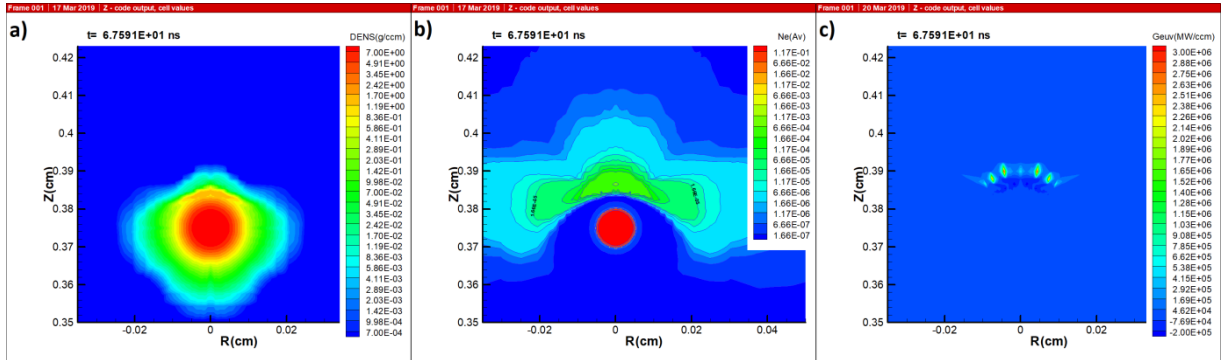

**Figure 10.** Iso-contour cross-sections for plasma mass density (left-hand side), electron density (middle) and in-band EUV emission power density (right) distributions at the in-band emission maximum time instant (t = 67.6 ns) for laser pulse of 150 mJ and 60 ns FWHM duration.

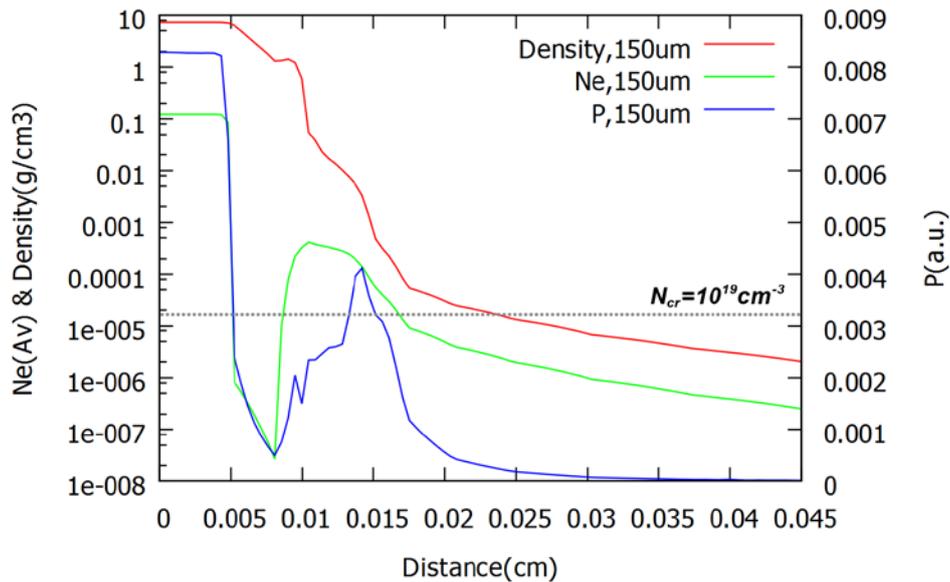

**Figure 11.** Calculated profiles of plasma mass- and electron-density (red and green lines, left-hand side log scale axis), and total plasma pressure (right-hand side linear axis) vs. distance from the Ø150 μm spherical target

centre, sliced along z-axis (r=0) at a time 67.6 ns for $CO_2$-laser pulse of 150 mJ energy and 60 ns time-duration. The grey dotted line displays the critical density value of the $CO_2$-laser light. Magenta background indicates an unstable zone, where $\nabla\rho\nabla P < 0$.

Laser light is absorbed mainly in the zone of critical electron density. The critical electron density for $CO_2$-laser infrared light with wavelength λ=10.6 µm is of $10^{19}$ cm$^{-3}$ = 1.66·$10^{-5}$ $N_A$. Mass density and electron density cross-section contours with labelled critical density value at a moment t=67.6 ns are shown in the Figure 10a) and b) respectively. Critical electron density is produced during laser light absorption and heating, ionization process is efficient in tin plasma at mass density of 3÷6·$10^{-4}$ g/cm$^3$ that is much less than liquid tin target density (7.28 g/cm$^{-3}$). The laser light doesn't penetrate deeper - closer to the target. The plasma flux from the target into critical density layer is provided by evaporation of target substance under irradiation by soft X-ray radiation from the plasma located above critical layer. If mass of the target is big enough to provide the plasma flux required (like in the case considered – the liquid tin droplet with ø150 µm) some kind of dynamic quasi-stationary plasma flux is formed. In this dynamic quasi-stationary plasma flux, an interlayer of relatively cold tin vapour with mass density of 1÷2 g/cm$^{-3}$ appears between the liquid tin droplet and low density plasma of the critical layer. Expanding of the tin vapour from the droplet provides the plasma flux to the critical layer. In critical layer the plasma is heated up and expands faster.

The plasma density in flux above the critical layer decreases to $N_e$=1.0÷1.6·$10^{-5}$ $N_A$, where the plasma electron temperature reaches the maximum of 124 eV (for laser intensity of 14 GW/cm$^2$), and expands further forming a corona. Plasma temperature in such corona decreases, as the cooling rate due to expansion and radiation cannot be compensated by laser power absorption in low density coronal plasma.

The optimal temperature for emission of in-band EUV photons of 92 eV energy is $T_e$=30÷40 eV. This temperature is attained in a zone below the critical layer at electron density $N_e$=7÷8·$10^{-5}$ $N_A$ - much higher of the critical one. The in-band EUV emission is localized in this zone (Figure 10c). It consists of bright spots because of thermal instability of the critical layer. Such non-uniform distribution of EUV emission under $CO_2$-laser irradiation was reported in previous our papers [17] and conference reports [28], also analogous calculated non-uniformities can be seen in [29] and the oscillations in EUV emission much deeper with respect to laser pulse modulation were observed in experiment [16]. There are various physical mechanisms of plasma instability at critical layer from Rayleigh-Taylor to thermal- type. The criterion for Rayleigh-Taylor (RT) instability in a smoothly stratified compressible medium is $\nabla\rho\nabla P < 0$ [30], i.e. if a gradient of mass-density ($\nabla\rho$) is opposite to the pressure gradient ($\nabla P$) [30]. This unstable situation always appears during the ablation of high density target because the produced plasma expands and drops more or less smoothly, but the plasma heating (the temperature maximum) is localized near the critical density. Especially, this effect becomes apparent for long wavelength laser ablation, where the critical density $N_{cr} \propto \lambda^{-2}$ is much lower of the target one. A typical distribution of plasma density and pressure of liquid tin target ablation under $CO_2$-laser irradiation is presented in Figure 11. Perturbations (as bubbles and spikes) of the critical layer caused by RT-instability allow the laser light to penetrate deeper and heat up plasma in bubbles increasing their development and producing non-uniformities in generated EUV emission as it is shown above. In many simulations by means of other codes, this instability of the critical layer was not observed in spite of valid criterion. A reason of that lies in enhanced numerical diffusion, especially if numerical scheme with Euler variables is used and the computational grid is not detailed enough.

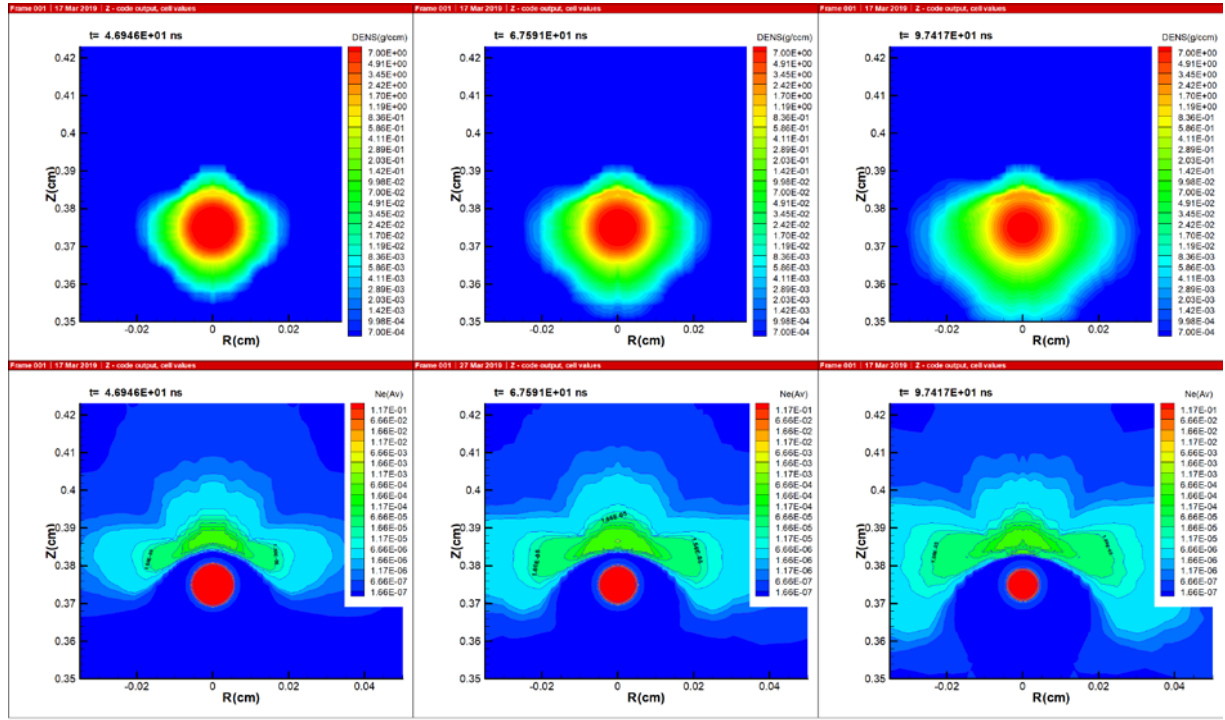

**Figure 12.** Plasma density (upper row) and free electron density (bottom row) dynamics as iso-contours of radial cross-section at three selected times: at t = 46.9 ns - before the EUV emission peak, t = 67.6 ns - at the peak and at t = 97.4 ns - after the peak. Electron densities are in Avogadro's number units. Iso-contour with critical electron density for $CO_2$ laser is labelled as "1.66E-5" (middle bottom image). Images for laser pulse of 150 mJ and 60 ns FWHM duration.

The temporal evolution of the plasma distribution at three time instants 46.95 ns, 67.59 ns and 97.42 ns is shown in Figure 12. During the time period of main contribution into in-band EUV output, the structure of layers of expanding plasma between the liquid target and corona conserves. These layers may be arranged into such series: a relatively cold dense layer beyond the liquid droplet, the layer of EUV in-band emission generation with $T_e=30 \div 40$ eV, the critical layer, the zone of temperature maximum just beyond it and, finally, the corona almost transparent for the laser light. This plasma structure is quasi-stationary and takes place at least for laser intensity of 14 GW/cm$^2$ amplitude with all of laser energies considered from 75 mJ to 600 mJ corresponding to laser pulse duration from 30 ns to 240 ns. The quasi-stationary structure explains the fact that in a rather wide range of $CO_2$ laser pulse durations and energies, the conversion efficiency (CE) of laser energy to in-band EUV output depends on laser intensity only. Almost identical results for in-band CEs in a wide range of $CO_2$ laser pulse durations were obtained experimentally in [16] for solid tin plate target. According to the simulation results, the angular distribution of in-band intensity, presented in Figure 13, almost all in-band radiation is localised in the solid angle less of $\pi$ steradian, therefore the CE of in-band radiation into adopted $2\pi$ is of 4% for 30ns FWHM and just slightly lower - of 3.67% for 240ns FWHM for equal laser intensities of 14 GW/cm$^2$ (Table 1). This slight decay of the in-band EUV yield with laser pulse duration is conditioned by an increasing of radiation re-absorption by expanding plasma from the target, as more cold plasma is produced with longer pulse. The concentration of in-band EUV emission in an annular cone between 15° and 45° to the laser beam axis allows the optimisation of the collector mirror shaping it as relatively narrow ring or using a grazing incident ellipsoid [31].

| Laser energy, mJ | Pulse duration, ns | In-band EUV, mJ | CE in 2π, % |
|---|---|---|---|
| 75 | 30 | 3.0 | 4.0 |
| 150 | 60 | 5.86 | 3.9 |
| 200 | 80 | 7.75 | 3.87 |
| 300 | 120 | 11.3 | 3.78 |
| 600 | 240 | 22.0 | 3.67 |

**Table 1.** EUV emission energy and CE in 2% spectral band at 13.5 nm wavelength as a function of laser energy and laser pulse duration.

In experimental measurements a pin-hole image of in-band plasma emission can be recorded using a CCD image camera (Figure 14). The Zstar RMHD code has a ray-tracing post-processing tool permitting to obtain such kind of image from simulated results for any given spectral band and pinhole position at specified time instant (a spectral intensity distribution) and a time-integrated image (a spectral fluence distribution) as well. In the Figure 15 an instant (at the maximum in-band emission t = 67.6 ns) and time-integrated raytraced pinhole images of the in-band EUV emission from tin plasma with $CO_2$-laser pulse 150 mJ / 60 ns are represented. Images generated through the pinhole located at 27 cm away from the target in lateral direction to incident laser beam (i.e. at 90º between laser beam direction and "source-pinhole" axe). Image angle is of 0.002º in X, Y axes (square matrix). The effects of radiation re-absorption are taken into account during the raytracing processing.

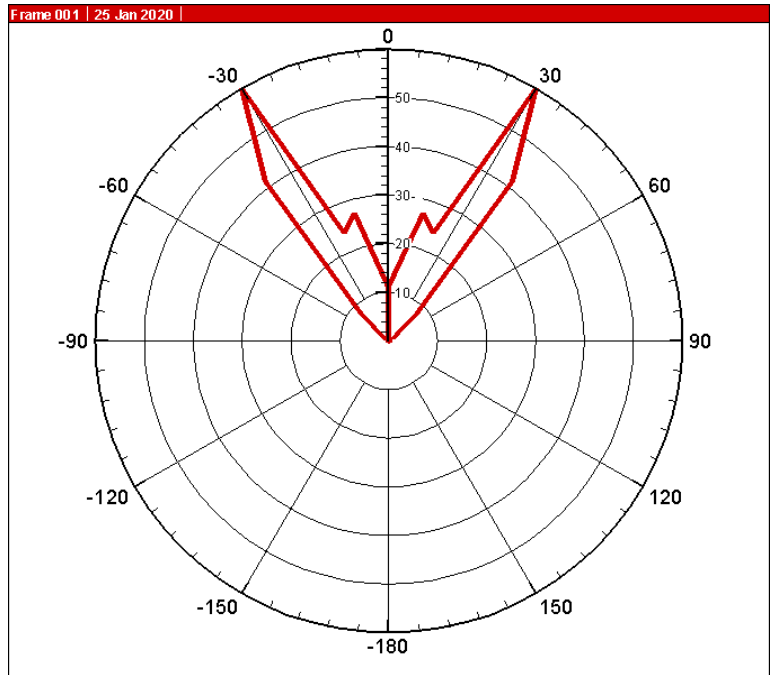

**Figure 13.** Polar diagram of intensity distribution $I\,(\theta)$ of in-band radiation group (in 2% waveband at λ=13.5nm) calculated for 30 ns laser pulse.

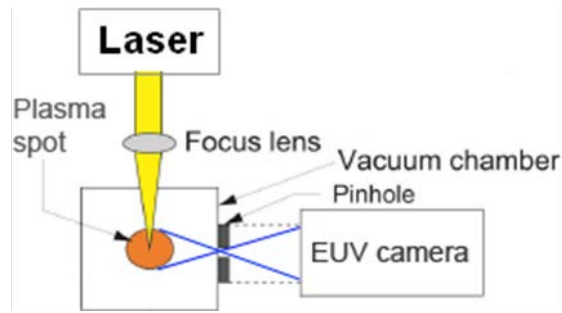

**Figure 14.** Scheme of experimental arrangement example for pin-hole imaging.

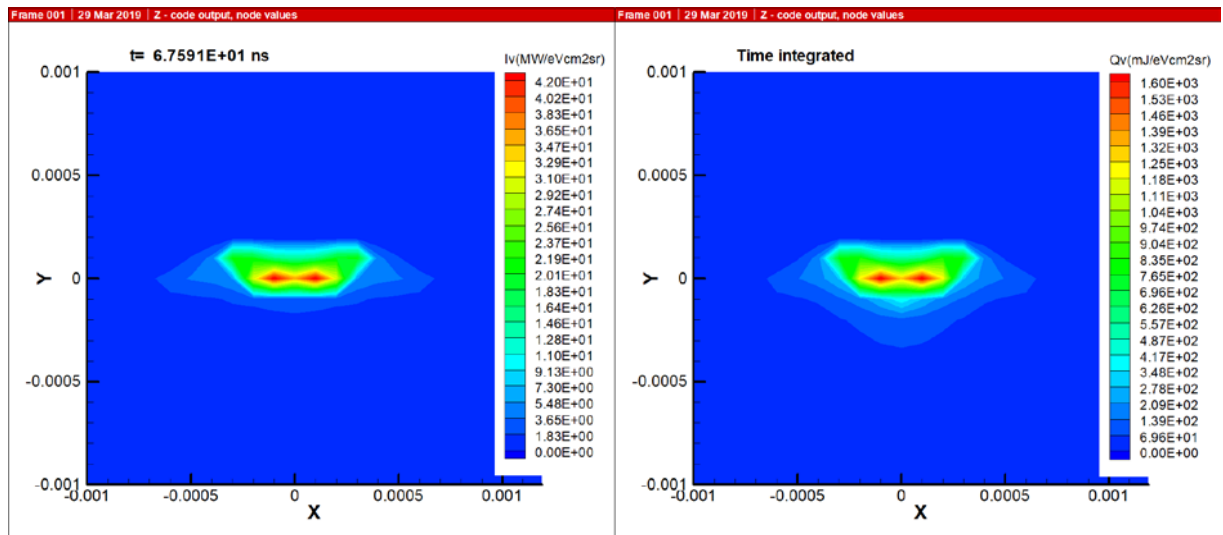

**Figure 15.** Instant (in EUV 2%-band of 13.5 nm at radiation peak, left-hand side image) and time-integrated (right) raytraced EUV images of the plasma source. Images are generated through the pinhole located at 27 cm away from the source in lateral direction to incident laser beam.

## 4. Conclusions

Calculations using the 2D RMHD code Zstar have shown the effect of pulse on the conversion efficiency and behavior of $CO_2$ laser-produced tin plasma. In the considered case the mass of the target is big enough to provide the plasma flux required to form a kind of dynamic quasi-stationary plasma flux. In this dynamic quasi-stationary plasma flux, an interlayer of relatively cold tin vapour with mass density of 1÷2 g/cm$^{-3}$ is formed between the liquid tin droplet and low density plasma of the critical layer. Expanding of the tin vapour from the droplet provides the plasma flux to the critical layer. The maximum conversion efficiency of 4% into adopted 2π was found using a 30 ns FWHM pulse and 75 mJ laser energy. The conversion efficiency decreases slowly with laser pulse duration increase for the same laser intensity of 14 GW/cm$^2$ and is about 3.67% for 240ns FWHM. This phenomenon is explained by the radiation re-absorption raise by expanding plasma, because the amount of colder plasma produced from the target is increasing with laser pulse duration. These results for in-band CEs correspond well to the values obtained before experimentally for $CO_2$ laser pulses with solid tin plate target. According to simulation results, the in-band EUV emission (corresponding to CE = 4%) is concentrated in an annular cone with less of 0.6π steradian solid angle allowing the optimization of a collector.

**Acknowledgments**
The authors would like to thank HUST "Second Home" Global Talents Recruitment Program and Basic and Applied Basic Research Major Program of Guangdong Province (No.2019B030302003) for funding provided.

**Data availability**

The data that support the findings of this study are available from the corresponding author upon reasonable request.

**Appendix**

**a. Plasma dynamics model**

The traditional ideal MHD model has been extended to take into account additional physics relevant to a EUV plasma source with realistic boundaries. It is necessary to account for the displacement current in the Maxwell equation to model the situation, when the conductivity tensor turns to zero, for instance, in not ionized neutral gas or insulator. A suitable physical model of the plasma includes a quasi-neutral plasma MHD description in a self-consistent electromagnetic field with ionization and radiation. In the MHD model a "zero mass" electron motion (generalized Ohm's law) with effective conductivity taking into account as well energy exchange between "electron" plasma component and the "heavy" (i.e. ions and neutrals) component, thermal conduction and radiation. The system of equations for the plasma mass density $\rho$, average plasma velocity $\vec{v}$, two temperatures $T_{e,i}$ in the self-consistent electromagnetic field $\vec{E}$, $\vec{B}$ in the RMHD approximation is (the Gaussian units are used)

$$\frac{\partial \rho}{\partial t} + \nabla(\rho \vec{v}) = 0; \quad \rho(\frac{\partial \vec{v}}{\partial t} + (\vec{v}\nabla)\vec{v}) = -\nabla p + \frac{1}{c}\vec{j}\times\vec{B} - \nabla \hat{\pi}_{ii}; \quad p = p_e + p_i$$

$$\nabla \times \vec{E} = -\frac{1}{c}\frac{\partial \vec{B}}{\partial t}; \quad \nabla \times \vec{H} = \frac{4\pi}{c}\vec{j} + \frac{1}{c}\frac{\partial \varepsilon \vec{E}}{\partial t}; \quad \nabla \vec{B} = 0; \quad \vec{B} = \mu\vec{H}$$

$$\vec{j} - \hat{\sigma}\frac{\vec{u}\times\vec{B}}{c} = \hat{\sigma}(\vec{E}^* + \frac{\nabla p_e}{en_e}); \quad \vec{E}^* = \vec{E} + \frac{1}{c}\vec{v}\times\vec{B}; \quad \vec{u} = -\frac{\vec{j}}{en_e} \quad (1)$$

$$\rho(\frac{\partial \varepsilon_e}{\partial t} + ((\vec{v}+\vec{u})\nabla)\varepsilon_e) = -p_e\nabla(\vec{v}+\vec{u}) - \nabla\vec{W}_e + \vec{j}\hat{\sigma}^{-1}\vec{j} + Q_{ei} - \nabla\vec{F}_r + G$$

$$\rho(\frac{\partial \varepsilon_i}{\partial t} + (\vec{v}\nabla)\varepsilon_i) = -p_i\nabla\vec{v} - \nabla\vec{W}_i - Q_{ei} - \hat{\pi}_{ii}\nabla\vec{v}; \quad W_{e,i} = -\hat{\kappa}_{e,i}\nabla T_{e,i}.$$

The radiation flux density $F_r$ in the electron energy equation is calculated by means of the radiation transport equation considered below. The term $G$ is introduced to take into account possible external heating, by laser light for example. The laser interaction with substance is considered separately below.

The quasi-neutrality of the plasma means $n_e \approx \bar{Z}n_i = \bar{Z}\rho/m_i$, because $\bar{Z}m_e/m_i \ll 1$. The Poisson equation is not introduced separately, so as to avoid introducing a new value as the charge density. It is fulfilled automatically from the second Maxwell equation, due to the fact that $\nabla\cdot(\nabla\times\vec{H}) = 0$, if initial conditions for the electric field satisfy the Poisson equation. In view of the special practical application of the physical equations we assume that at initial time, $\nabla \varepsilon \vec{E}\big|_{t=0} = 0$ everywhere in the space except, perhaps, at conductor boundaries.

The dielectric constant $\varepsilon$ and magnetic permeability $\mu$ are introduced to describe homogeneously the plasma dynamics together with the electromagnetic field behavior in insulators and metallic objects (if necessary). For the plasma and gas $\varepsilon = \mu = 1$ are imposed. Different plasmas and solid materials may be considered as differing from each other only by their properties, especially if solid materials can be sublimated and transformed to plasma medium. These properties of plasma and other substances are described by the equation of state (EOS), the thermal pressure $p_{e,i}(\rho, T_{e,i}, \bar{Z})$, specific internal energy $\varepsilon_{e,i}(\rho, T_{e,i}, \bar{Z})$ and ionization degree $\bar{Z}(\rho, T_{e,i}, U, t)$; the kinetic coefficients and tensors, i.e., electron–ion energy exchange rate $Q_{ei}(\rho, T_{e,i}, \bar{Z})$, electrical conductivity $\hat{\sigma}(\rho, T_e, \bar{Z}, \vec{B}, \vec{u})$, thermal conductivities $\hat{\kappa}_{e,i}(\rho, T_e, \bar{Z}, \vec{B})$, ion viscosity $\hat{\pi}_{ii}(\rho, T_e, \bar{Z}, \vec{B}, \vec{v})$, and radiation properties mentioned above and discussed below. Dependence of coefficients on the magnetic field is described in [20]. Here it is necessary to mention that, due to nonstationary effects, the ionization degree $\bar{Z}$ depends not only on the plasma density and temperature, but also on the time and radiation field, through the ionization kinetics described by level kinetic equations [17,21]. Dependence of the conductivities on the electron drift velocity $\vec{u}$ is described through the concept of the anomalous resistivity in the low-density plasma regions due to plasma turbulence: the low-hybrid drift and ion sound waves as having low thresholds.

The EOS and the kinetic coefficients are obtained by means of the cross-sections of collisional processes between the electrons, ions and atoms, and the interaction with radiation calculated in the frames of the general quantum-statistical model and ionization kinetics [17]. Plasma radiation properties, ionization and equation of state, as well as excitation and ionization rates, and plasma kinetic coefficients are calculated by means of interpolations from a set of tables prepared in pre-processing with the Hartree-Fock-Slater (HFS) model [21] in both the optically thick LTE, and the transparent non-LTE limits. The actual non-LTE condition at any instant is modeled by analytical interpolation between these two limits [19]. Interpolations of preliminary prepared databases allow the code to avoid on-line calculations of absolutely different processes like the plasma dynamics, the atomic physics and ion kinetics. As a result, the robustness of the code and accuracy of calculations of the main processes is enhanced considerably.

The computational model of equations (1) is developed on a numerically stable completely conservative, implicit difference scheme in Euler-Lagrange variables. The Euler-Lagrange variables retain the advantages of the Lagrange and the Euler variables reducing the numerical diffusion (with respect to pure Euler variables) and avoiding grid crossing. The solution of the MHD equations using implicit schemes provides a better precision at larger time step. A completely conservative scheme is important for the simulation of a physical system, as it ensures that the solutions obtained are physical, since all integrals of the system are fulfilled at each time moment.

**b. Radiation transport**

The influence of radiation on spectral characteristics of plasma and ionization degree is important. Partial radiation trapping can induce a significant deviation from the optically thin coronal equilibrium or collisional-radiative equilibrium (CRE) approximations for non-LTE plasma description as well as from LTE. Meanwhile, at high temperatures the radiation field is formed in the process of plasma dynamics, affecting the microstates of the plasma ions, and the radiative properties of plasma. Respectively, the microstates of the ions determine the emissivity and spectral absorption coefficients of the plasma. Re-absorption in the spectral lines can change the ionization stage and internal energy of the plasma at the same electron temperature, as these are strongly coupled to the electrons via level kinetics.

The radiation transfer equation for spectral intensity $I_\omega$ with photon energy $\hbar\omega$, without free electron scattering processes, has the form

$$\frac{1}{c}\frac{\partial I_\omega}{\partial t} + (\vec{\Omega}\nabla)I_\omega = j_\omega - k_\omega I_\omega, \quad (2)$$

where $c$ is a speed of light, $\vec{\Omega}$ is the unit vector in the direction of the radiation ray. The spectral absorption coefficient $k_\omega$ and emissivity $j_\omega$ are written in a form taking into account an induced emission. They depend on the local properties of a plasma, e.g., density and temperature and the ion distribution, which, in turns, are dependent on the radiation field and should be calculated in non-LTE, i.e., depending on spectral radiation field through the level occupancy balance considered in the last section. From the radiation transport equation (2) radiation energy density $U = c^{-1}\iint I_\omega d\vec{\Omega} d\omega$ and flux $\vec{F}_r = \int \vec{F}_\omega d\omega = \int \vec{\Omega} I_\omega d\omega d\vec{\Omega}$ are calculated and used in the energy balance of the RMHD system of equations (1).

As a rule the characteristic size of the DPP or LPP plasma of interest is much smaller than the distance traversed by the radiation during the evolution time, and quasi-stationary approximation may be applied, i.e., the time derivative in eq. (2) may be neglected. To find the radiation field in the quasi-stationary case, by integrating the radiation transport equation along the trajectories under cylindrical symmetry conditions, we obtain an expression for the intensity [22]:

$$I_\omega(r,z,\phi,\theta) = \int_0^\tau \frac{j_\omega}{\kappa_\omega} e^{\tau'-\tau} d\tau,$$

where $r, z$ are the radial and axial coordinates; $\varphi, \theta$ are the spherical angle coordinates of the trajectory: $\varphi$ between trajectory projection onto a plane perpendicular to the Z axis and the radial direction, and $\theta$ between trajectory and the Z axis; $\tau = \tau(x) = \int_0^x \frac{\kappa_\omega(r,z)}{\sin\theta} dx$ is the optical depth, $\tau' = \tau(x')$. The coordinate $x = x(r,\varphi)$ is taken along the projection of the ray onto a plane perpendicular to the axis Z.

**c. Laser energy transport**

The laser light transport is calculated by means of simplified two-direction transfer model [23] taking into account an absorption and reflection of the laser light along arbitrary given trajectory $\vec{s}(r,z)$, i.e.,

$$\frac{\partial \psi_+}{\partial \vec{s}} = -(\kappa_l + r_l)\psi_+, \quad -\frac{\partial \psi_-}{\partial \vec{s}} = -\kappa_l \psi_- + r_l \psi_+,$$

where $\psi_+$, $\psi_-$ are flux densities of direct and reflected laser light respectively, $\psi = \psi_+ + \psi_-$. The reflection coefficient $r_l$ from the critical density zone $\omega_{pe}^2 > \omega_l^2$ is $r_l = \sqrt{\omega_{pe}^2 - \omega_l^2}/c$. The laser light absorption coefficient $\kappa_l = \kappa_{st} + \kappa_r + \kappa_{bi} + \gamma\kappa'_{bb}$ includes an interaction of radiation with electrons, ions and neutral atoms: $\kappa_{st}$ for collisional (inverse-bremsstrahlung) absorption with plasma dispersion properties [24], $\kappa_r$ resonant absorption (a critical density effect due to Longmuir plasma oscillation resonant excitation) [24], the effective bond-bond excitation calculated from spectral tables at laser quantum energy $\hbar\omega_l$ with taking into account the induced deexcitation, $\kappa_{bi}$ direct (if laser quantum energy is higher than the ground state ionization energy) or tunnel (if laser quantum energy is lower than the ground state ionization energy) ionization calculated from spectral tables also; the parameter $\gamma$ is introduced to take into account a probability of dissipation of absorbed laser energy in bond-bond electron excitation to thermal energy. The conversion of laser energy absorbed due to level excitation to plasma realizes through non-radiating deexcitation described by probability $w_{3b}$. The stepwise ionization described by probability $w_n$ is also possible if several quanta are absorbed consequently, i.e. the first quantum excites a ground state, the second one excites once more, etc and just the last one ionizes the atom. The parameter $\gamma$ is defined by these probabilities and spontaneous emission probability $w_{se}$, viz. $\gamma \approx (w_{3b} + w_n)/w_{3b} + w_n + w_{se}$, calculated from spectral tables. Under these approximations, heating of electrons by laser light flux $L_e$ in the energy equation (1) is

$$G = \nabla \vec{L}_e = \kappa_l \psi .$$